# Using LLMs to Enhance Democracy

Seth Lazar (ANU, ORCID iD: 0000-0001-5378-6033)

Lorenzo Manuali (University of Michigan, ORCID iD: 0009-0006-3215-5703)[*]

Machine Intelligence and Normative Theory (MINT) Lab

## Abstract

LLMs are among the most advanced tools ever devised for understanding and generating natural language. Democratic deliberation and decision-making involve, at several distinct stages, the production and comprehension of language. So it is natural to ask whether our best linguistic tools might prove instrumental to one of our most important tasks involving language. Researchers and practitioners have recently asked whether LLMs can support democratic deliberation by leveraging abilities to summarise content, to aggregate opinion over summarised content, and to represent voters by predicting their preferences over unseen choices. In this paper, we assess whether using LLMs to perform these and related functions really advances the democratic values behind these experiments. We suggest that the record is mixed. In the presence of background inequality of power and resources, as well as deep moral and political disagreement, we should not use LLMs to automate non-instrumentally valuable components of the democratic process, nor be tempted to supplant fair and transparent decision-making procedures that are practically necessary to reconcile competing interests and values. However, while LLMs should be kept well clear of formal democratic decision-making processes, we think they can instead strengthen the informal public sphere—the arena that mediates between democratic governments and the polities that they serve, in which political communities seek information, form civic publics, and hold their leaders to account.

---

[*] Equal contribution. This work is supported by Australian Research Council grants FT210100724 and LP210200818. It is also supported by Institute for Humane Studies under grant no. IHS018415. We thank the members of the MINT Lab, as well as Ricardo Fabrino and John Dryzek for comments on earlier drafts. We also thank the organizers and participants of the Institute for Humane Studies AI and Cybersecurity workshop for comments and the opportunity to present this work. This is a preprint of a work in progress, accepted as a non-archival submission to ACM FAccT 2025. Comments welcome to seth.lazar[at]anu.edu.au.

# 1. Introduction

Following the hubris of the 1990s (Fukuyama, 2020), the last two decades have seen democratic institutions come under increasing pressure, with the global rise of authoritarianism and a growing sense that democracy is in crisis (Diamond, 2020; Kingzette et al., 2021; Scheppele, 2022; Grumbach, 2023). While much of this concern has been driven by the concerted efforts of autocrats (and would-be autocrats) to use legalistic means to undermine democratic freedoms (Huq and Ginsburg, 2018; Scheppele, 2018), it also owes something to our growing collective dependence on digital technologies controlled by powerful technology corporations (Lehdonvirta, 2022; Susskind, 2022; Lazar, 2025). In early 2023, GPT-4's launch announced the arrival of highly capable large language models (LLMs) into this roiling stew, and they immediately induced further anxiety among democracy's partisans (Kreps and Kriner, 2023; Allen and Weyl, 2024; Coeckelbergh, 2024). Existing narrow AI, for all its flaws and inefficiencies, had nonetheless already amplified the power of a few corporations over vast swathes of our lives (Crawford, 2021; Lazar, 2024b). It would be natural to assume that the advent of much more capable AI systems would exacerbate this trend.

It is therefore unsurprising that those who care about both democracy and fostering the advancement of AI research would explore whether AI might fix what it threatens to break (Ovadya, 2023; Siddarth, 2024; Weyl et al., 2024). In a number of research papers, computer scientists and AI engineers have sought to use LLMs to enhance democratic decision-making.[1]

Our aim here is to explore whether and to what extent using LLMs to support democratic deliberation and decision-making actually serves democratic goals. Given that many of these projects were funded by the very Big Tech companies that some accuse of destabilising democracy, some might dismiss such initiatives as mere 'democracy-washing' (Himmelreich, 2023). However, we take these research projects on their own terms, instead of further discussing their motivations or funding.

In the next section, we introduce our approach and classify the different ways in which engineers have used LLMs to enhance democratic decision-making and deliberation. In Section 3, we identify six different non-instrumental and instrumental values that democratic institutions serve. In Section 4, we ask how well different approaches to LLM-enhanced democracy serve those values. In Section 5 we seek to explain the shortcomings identified in section 4, by identifying democratic functions that LLMs cannot, in principle, serve—as well as those that they can. Section 6 concludes.

---

[1] Our survey of the literature was conducted in mid-2024 (note that some papers were preprinted before their final publication date). We used scholarly databases and search tools to identify papers seeking to use LLMs to support democratic deliberation or decision-making. We have monitored the field since conducting our initial survey and have confirmed that newer papers continue to fall into the taxonomy that we inferred from the existing literature. The key papers discussed here are: Bakker et al., 2022; Anonymous, 2023; Argyle et al., 2023; Devine et al., 2023; Edelman, 2023; Edelman and Klingefjord, 2023; Konya et al., 2023; Landemore, 2023; Mendoza, 2023; Small et al., 2023; Zaremba et al., 2023; Gudiño-Rosero et al., 2024; Huang et al., 2024; Tessler et al., 2024; Fish et al., 2025; Konya et al., 2025. We also note in passing some applied projects, such as Meta's use of deliberative polls to devise rules for governing bullying and harassment in virtual reality spaces (Sulots, 2023). For an overview and living history of some of these initiatives, see Goldberg et al., 2024.

## 2. LLMs for Democracy

The intersection between democracy and digital technologies has been studied by philosophers and social scientists for decades.[2] Some scholarship argues that digital technologies threaten democratic norms and institutions—for example due to the concentration of power in tech companies, or the ways in which social media have undermined the digital public sphere.[3] Some proposes radical approaches to taking back democratic control of digital industries and artefacts.[4] Some criticises the co-optation of the ideology of democracy by a tech industry that conflates mass distribution with democratisation.[5] And some imagines a digital future where AI and related technologies enable completely novel forms of democratic engagement.[6] Similarly-motivated computer scientists and AI engineers have approached these problems from the other direction—aiming to build the future of democracy and AI, not assess it. But until recently they had to mostly confine themselves to narrow exercises in computational social choice and bold proposals for new kinds of processes that further advances in AI might make feasible.[7] That changed in early 2023.

Large language models are neural networks pretrained using self-supervised learning on vast quantities of text converted into sequences of numerical tokens. The objective function in pretraining is to predict masked tokens (in GPT models like GPT-4, to predict the next token in a sequence). They are then fine-tuned using supervised learning and reinforcement learning to perform more complex functions, for example text summarisation or dialogue assistance (being a good chatbot), while also abiding by behavioural norms and constraints.[8] Since mid-2024, the most capable current LLMs are further trained with reinforcement learning to make use of additional computational resources at 'inference time' (for example, when responding to a prompt). Unlike earlier forms of AI, LLMs are truly general purpose—while the pretrained models might be described as 'autocomplete on steroids', fine-tuning unlocks latent capabilities in many different domains. It is important to be extremely clear about three points. First: LLMs constitute a step-change in what is feasible with AI. They are radically different in this respect from the AI technologies that preceded them. Second: Their capabilities are consistently and rapidly improving. Third: they open up the possibility of building with AI to anyone, not just engineers.

Democracies depend on deliberation and collective decision-making. Deliberation can occur in the formal setting of a citizens' assembly, or the informal setting of a pub conversation the night before an election, and everything in between.[9] The key moment in collective decision-making is when we vote for the holders of elected office. But there are many other moments of collective decision-making in a well-functioning democracy, from town hall meetings to nationwide referendums.

---

[2] For example, see Wiener, 1950; Mcluhan, 1964; Mumford, 1964; Westin, 1971; Winner, 1984; Sterling, 1986; Arterton, 1987; Bjerknes et al., 1987; Ronfeldt, 1992; Bonchek, 1995; Groper, 1996; Buchstein, 1997; Margolis et al., 1997; Hill and Hughes, 1998; Wright, 2018; Bernholz et al., 2021; Coeckelbergh, 2024.

[3] Khan, 2016; Theocharis et al., 2016; Brady et al., 2017; Wu, 2018, 2019; Cohen and Fung, 2021; Reich et al., 2021; Siegel, 2022; Susskind, 2022; Aytac, 2024.

[4] Muldoon, 2022; Fischli, 2024

[5] Sætra et al., 2022; Himmelreich, 2023; Seger et al., 2023; Lin, 2024

[6] Bernholz, 2016; Gastil and Davies, 2020; Schuler, 2020; Bernholz *et al.*, 2021; Landemore, 2021b; Himmelreich, 2022; Fischli, 2024; Grossi et al., 2024

[7] Brill, 2019; Lee et al., 2019; Benade et al., 2021; Brill, 2021; Flanigan et al., 2021; Ford, 2021; Kahng et al., 2021; Landemore, 2021b

[8] For good overviews of how LLMs work, see Naveed et al., 2023; Millière and Buckner, 2024. Of course, while LLMs are presently at the frontier of AI research, other new advances could supersede them—whether our argument here transfers will depend on the specific technical details.

Democratic deliberation and decision-making both involve, at several stages, the production and analysis of language. LLMs are among the most advanced technologies ever devised for analysing and generating linguistic content.[10] So it is natural to ask whether our best and most adaptable tools for manipulating language might prove instrumental to one of the most important ways in which we use language. Moreover, because LLMs are so capable and so easy to deploy, over the last two years AI engineers, computer scientists, and even hobbyists have sought to build applications using LLMs for practically every conceivable function. We therefore find ourselves at a rare moment in history, in which one of the most productive tasks for political philosophers and theorists is not just to evaluate technology's impact on democracy through the lens of a bystander, but to actually *build* democratic technologies. As a complement, we are witnessing the growth of a new subfield—what one could call, by analogy with computational social choice, computational political theory.

We collected and reviewed this growing literature on using LLMs to enhance democratic deliberation and decision-making—focusing on papers published in the year or so after GPT-4's release. From this literature, we identified four distinct roles with which LLMs have been tasked.

First, researchers and practitioners have asked whether LLMs can support democratic politics by leveraging their *summarisation* capabilities. Second, they have explored how they can support democratic decision-making by *aggregating* opinion over options, to find consensus or to optimise for some metric of collective support. Third, some projects use LLMs to *represent* individuals in a further democratic decision procedure (in general, by predicting their judgments over policy propositions). Finally, fourth, some have also explored whether LLMs can *facilitate* deliberation. We next elaborate on each of these approaches in turn.

## Summarisation

The prospect of using LLMs to summarise public submissions to government consultations has been more widely-touted by ministers than by researchers, and while several active projects are exploring whether and how this can succeed (Small *et al.*, 2023; Straub et al., 2024), few results have been published so far.[11] The basic theory, however, is very simple: public consultation is a key input to democratic decision-making; if consultations are to be inclusive they will involve the production of a vast amount of text;

---

[9] For examples of work that discuss varying sites of democratic deliberation, see Mansbridge, 1983; Habermas, 1991; Schmitt-Beck and Grill, 2020.

[10] No other technique has come close to LLMs' fluency in generating linguistic content. And though, of course, statistical methods for linguistic analysis are undoubtedly invaluable, LLMs demonstrate a helpful understanding of linguistic structure that gives them the potential to advance the field of linguistics (Beguš et al., 2023; Millière, 2024).

[11] For a very early discussion of this approach, pre-AI, see Schlosberg et al., 2008. Johannes Himmelreich (2022) also moots the possibility of using AI to 'identify citizen input during public consultations as novel, detailed, or otherwise relevant'.

democratic governments, therefore, must process that text and extract insights from it. This is extremely labour-intensive; civil servant analysis of these documents is also (plausibly) lossy and biased, simply in virtue of our inability to reliably process very large amounts of information (Bennett and Cutler, 2019). LLMs have been trained for the task of summarisation (Ouyang et al., 2022) and in general perform well at it (Zhang et al., 2024). It is an obvious thought, then, that we can enhance government responsiveness to public consultations by extracting insights using LLMs.

This basic capability of LLMs has been used more creatively in the 'Recursive Public' project, which built on vTaiwan, the scheme for channelling public input into Taiwan's political system led by Audrey Tang, until recently Taiwan's digital minister (Devine *et al.*, 2023). The project uses Polis—an online platform that allows users to make short statements and to indicate their support (anonymously) for statements made by others—to gather participants' inputs to set the agenda for policy discussion concerning AI. Then they used AI in two ways to summarise those discussions—first, they used an LLM to summarise key arguments and points from the discussion, and to cluster similar arguments and topics; then they used a further tool ('Talk to the City') to convert that data into a 'a two dimensional space where semantically similar arguments are positioned close to each other' (Devine *et al.*, 2023: 20). The end result of this deliberative agenda-setting exercise was a list of priorities concerning AI. They then did a deep-dive deliberation on one of these priorities using the same deliberative-and-AI-summarization process, with the end result a set of guidelines for policy on that topic.

## Aggregation

Summarisation involves selection and aggregation—similar views are clumped together. At the limit, summarisation is a decision procedure whereby a diverse array of inputs are converted into a single output. Several recent research projects have explored using LLMs in deliberation to identify a larger or smaller set of statements which to some extent represent the views of the deliberators. In these examples, the LLM plays the role in relation to text-based descriptions of preferences and values that an aggregation function plays in relation to numerically-represented preferences in social choice theory. Notably, however, where social choice theory can only determine support for a fixed list of policy positions over which participants have expressed preferences, LLMs can remove this constraint, allowing us not only to identify statements that receive optimal support against some measure, but to generate new statements that will be even more widely endorsed.

This is most rigorously explored in 'Generative Social Choice' (Fish *et al.*, 2025).[12] Focusing on participants' preferences over norms for personalising chatbots, the authors first elicit statements from the participants about what those norms should be. These are the options. They also gather participants' preferences over those options. Classical social choice would stop here, and aggregate, according to some function, their preferences over those options to identify which if any were socially optimal. Generative social choice goes further. First, it elicits further information from participants about their values and opinions. Then they use all of these inputs—statements, preferences, and additional

[12] For other related examples, see Bakker *et al.*, 2022 and more recently Tessler *et al.*, 2024.

data—as material with which to prompt GPT-4 to propose *new* statements that it predicts will attract more support than those that are extant, and in addition to 'act as a proxy for the participant, predicting their preferences over any alternative, whether foreseen or newly generated' (Fish *et al.*, 2025: 2). The authors show that their approach can be used to generate new statements that better fulfil quantitative criteria of representativeness than the existing statements.[13] They also validate the predictions of people's preferences by exposing participants to the slate of generated statements after the fact, finding that '84% of participants say that their assigned statement captures their view on abortion policy "excellently" or "exceptionally"' (Fish *et al.*, 2025: 4).

Where 'Generative Social Choice' uses LLMs to help find consensus among disaggregated participants who engage with each other only asynchronously, 'Democratic Policy Development Using Collective Dialogues and AI' (Konya *et al.*, 2023) involves live deliberation but ultimately uses LLMs for a similar purpose. In this project, researchers invited participants to first read a discussion guide prepared by the facilitation team, and then to engage with one another on Remesh, another text-based online deliberation platform. They also made statements, proposed policies, and evaluated each other's statements and opinions via pairwise ranking and binary agree/disagree voting. These rankings over the subset of statements that a user sees are then used to extrapolate their preferences over all other statements. Then the LLM is called to create a summary of the statements that exhibit the lowest amount of disagreement ('bridging responses'); this is further used as an input for a new set of policy proposals, which also draw on the list of bridging responses and two further examples. The resulting policies are then given 'justification scores', reflecting how close they are to the bridging responses, through combining semantic similarity measurement and the LLMs' own quantitative similarity rating. Again, the role of the LLM is to not just summarise and select among the statements that were actually produced by the participants, but to use those statements (and other data from the deliberation) as inputs to generate statements that (according to some measure) are preferable to those that participants actually proposed.[14]

## Representation

A number of papers rely at some point on using LLMs to *represent* individuals by predicting their preferences over unseen policy options. This includes, tangentially, 'Generative Social Choice' and 'Democratic Policy Development'. But the approach is much more central in others, which explicitly consider whether we can train LLMs to be representatives, proxy agents acting on a principal's behalf in a democratic process. In 'LLMs as Agents for Augmented Democracy', the authors try to train LLMs to predict an individual's political preferences (using survey data from the 2022 Brazilian election), with a view to those LLMs then standing in for the person they represent in a collective decision-making procedure (Gudiño-Rosero *et al.*, 2024). The same approach is taken in 'Language agents as digital representatives in collective decision-making' (Jarrett et al., 2025). Whereas the other examples in this category involve LLMs facilitating judgment aggregation, this approach uses the LLM to represent individual agents' inputs into the final decision procedure.

[13] Their principal representation criterion, Balanced Justified Representation, is informally described as follows: '[I]f there is a coalition of agents that is (i) large enough to "deserve" a statement on the slate by proportionality and (ii) has cohesive preferences (i.e., there is a statement for which all these agents have utility at least θ), then (iii) the coalition must not be "ignored"' (Fish *et al.*, 2025: 6).

[14] Unlike with 'Generative Social Choice', however, this wasn't the final step. Next, the facilitation team selects a further subset of policy responses to form an initial slate of policies, which is elaborated on by experts chosen by the facilitation team, to refine the policies for clarity and to capture edge cases. Then participant input is used to refine the policies by giving reasons for and against supporting given policies in the slate. The policies are then refined in light of that feedback to make them more representative. Finally, a third phase is run such that the public can deliberate and vote on each policy and the slate as a whole (Konya *et al.*, 2023).

### Facilitation

The foregoing methods involve either human-mediated deliberation, or asynchronous collection of statements and opinions from participants. Other projects also sought to use LLMs to *facilitate* the process of deliberation itself. Rappler's initiative (Mendoza, 2023) involved creating an AI-moderated chatroom (called aiDialogue), in which participants in focus groups chat with an LLM that asks them questions, including follow-ups, and then generates summaries from their inputs. Then, as with the other projects above, the LLM is prompted to generate policy ideas based on the conversation and summaries, which participants are asked to vote on individually. 'Common Ground' (Anonymous, 2023) used LLMs to seed and moderate synchronous deliberation among participants, as a means of keeping discussion on-topic and also moderating the conversation.[15]

'Democratic Fine-Tuning' (Edelman, 2023; Edelman and Klingefjord, 2023) takes a similar, albeit less social approach. Participants engage in 1:1 deliberation with an LLM-based chatbot about a given topic. The chatbot prompts them to give information about the values they think are at stake concerning the given topic. It then synthesizes the conversation into a few key values that (in its view) the participant considers important (notice that this involves some very lossy compression—transforming whole conversations into a few characteristic values).[16]

## 3. What is the Point of Democracy?

Political theorists are rarely enough asked to contribute even to writing the laws on which new political organisations are based; the opportunity to build new democratic technologies is hard to pass up. But, while the initiatives described in the previous section all have clearly democratic motivations, how can we assess whether and to what extent they really serve democratic values? What, indeed, *are* democratic values? Obviously we cannot settle this question in the space available; however, we highlight six key values, widely recognised in the literature on democratic theory, which can provide a barometer against which to measure these initiatives.[17] For ease of exposition, we will distinguish between non-instrumental and instrumental values (Anderson, 2009).[18]

---

[15] Again, more recently see Tessler *et al.*, 2024.
[16] From that point, the approach implements an ethical theory according to which our objective should be to act according to the 'wisest' values, where value A is wiser than value B if A 'contains' B. They then generate a 'moral graph' depicting these relationships, which can in principle be used for fine-tuning other LLMs.
[17] We lack space to defend these values in this paper; however, the cited works include many robust defences of each value that we invoke.
[18] Anderson, 2009 position is that democracy's non-instrumental value is *conditional on* its instrumental value. In a sense, this renders the distinction less sharp than we make it out to be. But, for the sake of simplicity and clarity, we think it offers a nice demarcation.

The pragmatics of politics and the societal division of labour mean that *actual* influence over societal outcomes can never be equally distributed. Some people seek and hold positions of executive authority that inescapably place them, in that respect, 'above' others. Democratic decision procedures affirm in spite of these pragmatic realities the equal standing of all those who have a say (Dahl, 1989). They establish that all exercises of political power—all uses of the resources and normative authority of the state—are ultimately subject to the control of a decision-making mechanism into which every democratic citizen has a right to contribute equally (Allen, 2023; Kolodny, 2023).

This control matters not only because it affirms the basic equality of all those who contribute to it. It is also constitutive of *accountability* for particular decisions, and authorisation for the broader exercise of power by those who act on our behalf (Christiano, 2004; Viehoff, 2014). This is both a necessary kind of counter-power that counterbalances the power those who act on our behalf have over us (Pettit, 2012), and the means by which we collectively exert positive control over the institutions that serve us (Miller, 1995; Lazar, 2025).

The third key non-instrumental value of democratic politics is the opportunity it provides for the exercise of civic virtue through participation in political life (Macedo, 1990). This notion has deep roots in Aristotelian political philosophy (Winthrop, 1978), but it was repurposed for the era of political liberalism in the 1990s (Macedo, 1990), and in addition has republican adherents (Dagger, 1997). On these views, active political participation is a non-instrumental good—part of what it means for a life to go well—and democratic institutions are necessary for this to be a realisable good for everyone.

Turn next to democracy's pragmatic virtues—the goods it helps us to realise. First is the hard-nosed advantage it provides to the realisation of stable government in the presence of divergent interests (Lijphart, 1977; Miller, 1983). This is in part because transparent democratic procedures are the best means available for providing those who lose a political contest with clear and obviously justified reasons for them to accept defeat (Anderson, 2005). This is true even in regular collective decision-making. Everyone can understand the legitimacy of majority vote, and if you have just seen the procedure play out, and seen that you lost, then that gives you a strong (though obviously not decisive, especially if you are in a persistent minority) reason to accept the outcome. But democracy is also an excellent means of identifying possible compromises between competing interests. By creating scope for veto players in collective decision-making (Tsebelis, 1999), democratic procedures induce us to search for and identify compromise positions that can evade those vetoes.

As well as enabling us to achieve a *modus vivendi* (Rawls, 1993), democracy can be the imperfect means whereby we approach something that better approximates a general will (Rousseau, 1762), or at the very least gives us the opportunity to change our political preferences through interacting with one another (Young, 2002). When the democratic process enables us through engagement with the other to actually move closer to them in our values and political preferences, then this *transformation* not only quells instability and unrest, it enables a genuine form of collective self-determination (Lovett and Zuehl, 2022).

Finally, one of the most popular arguments for democracy is that it enables us to make better decisions—it draws on our collective intelligence to make epistemically better choices (Landemore, 2012a). Some arguments to this end are descriptive—showing that democracies have in fact yielded more stable and prosperous societies (Quinn and Woolley, 2001). Others are pragmatic, asserting that societies should aim at justice, and only political institutions that empower everyone to assert their interests will realise that goal (Allen, 2023). Others focus on theoretical results, such as the Condorcet theorem, and broader work on the wisdom of crowds (Cohen, 1986; Goodin and Spiekermann, 2018).

For brevity, we will refer to these values respectively as *equality*, *authorisation*, *participation*, *conciliation*, *transformation* and *collective intelligence*. Now, obviously these values matter *much more* in the kinds of high-stakes decisions for which we typically mobilise democratic procedures. We should care much more about equality and authorisation, for example, when thinking about the deployment of the coercive power of the state than when pondering conversational norms for chatbots. However, each of the papers discussed in Section 2 is offered in the democratic spirit, and so it is fair to consider how well they might realise these six goods.

## 4. Evaluating Democratic LLMs

We now have a set of four democratic functions to which LLMs have been put, and six democratic values. The next step is to evaluate the technologies described in section 2, to see how well they realise those values. We again divide our analysis according to those four functions.

### Summarisation

Using LLMs to summarise responses to public consultations, or long and complex deliberations, or even policy debates more generally, is perfectly consistent with the non-instrumental goods of democracy, and perhaps positively supports participation and authorisation because it promises to lower the burdens to participation, and could potentially thereby also provide for more robust accountability. Modern communication technologies have made it extremely easy for political institutions to solicit input from the public, but the sheer volume of that input is hard to effectively process. Better means to that end are clearly a plus. And the same basic analysis works for the instrumental goods of democracy—more efficient information processing should support conciliation, as well as the transformation of preferences and ultimately the surfacing of a wider variety of views, supporting collective intelligence. This is simply because each of these goods is advanced by creating better systems for collecting and interpreting public inputs to consultations. If you believe that you had a real opportunity to contribute to shaping public policy, and if you can see the reasons given by those who took the other side of a debate, then that should make you more inclined to accept an outcome that is not in your favour. If 'public opinion' is being accurately recorded and processed, then we should be better able to see what the 'general will' is. And if we have efficient and effective means for processing these public inputs, we can better identify and make use of the wisdom of the crowd.

However, the foregoing is all plausible only if we are confident that the summarisation process is neither excessively lossy, nor ideologically skewed, nor indeed hallucinatory. If instead the use of LLMs to summarise collective inputs, or deliberations, or individual conversations with a chatbot are inadequately tethered to their source material, then that is clearly a problem—it might undermine equality insofar as some voices might be amplified above others; it threatens authorisation if key details are missed or hallucinated; civic virtue is hardly enhanced by participating in a charade; and inaccurate summaries cannot support conciliation (why should we accept the outcome of a flawed process?), preference transformation, or collective intelligence (except accidentally).

The problem is that LLMs' record at summarisation is somewhat mixed. They excel at some tasks, such as summarising short news articles (Zhang *et al.*, 2024). However, they perform less well when summarising long inputs. They both have a propensity to make things up (Kim et al., 2024), and do a poor job of integrating insights that are widely distributed around the input source (Cheng-Ping et al., 2024). For example, researchers used LLMs to attempt a summarisation and analysis of an NTIA call for public comments, and found that the LLMs would often introduce subtle but significant changes of meaning (Boulos et al., 2023). LLMs also display well-documented political biases, which are likely to interfere with a summarisation that should be politically neutral (Liu et al., 2022; Santurkar et al., 2023; Taubenfeld et al., 2024; Rettenberger et al., 2025). Based on work concerning "generative monoculture", there's evidence to suggest that LLMs' summaries too often do a decent job of summarising the highest frequency but lowest signal content from a dataset, but perform much worse at drawing out high signal, low frequency content (Wu et al., 2024). That is, when summarising text inputs they have trouble picking out outliers in data that could convey potentially important information. This might mean failing to notice the one or two brilliant ideas in the public consultation, undermining collective intelligence. Or it could mean ignoring the minority of voices raising deep concern about how a proposed policy will seriously harm them, thus undermining equality and conciliation.

Of course, the humans that currently summarise responses to public consultations also have many cognitive and moral biases. And LLMs hardly need to be perfect to constitute an improvement. Our claim here is not that LLMs *could never* be competent enough summarisers for this function to contribute valuably to democratic decision-making. But we should expect a higher standard from a computational means of democratic information processing than from humans. Human summarisers can be held accountable for their errors, and their biases are likely to cancel one another out over enough trials due to any given individual's limited ability to process information. LLMs for summarisation obviously cannot be held accountable, moreover they can be applied to vastly more cases, and as such their biases can ramify much more widely (Bommasani et al., 2022). In addition, our ability to intervene directly on LLMs' summarisation abilities should allow us to aim higher.

Moreover, the deeper problem is that we are not at present measuring the right things to know whether LLMs can play this information processing role in democracy.[19] The papers

---

[19] For example, summarisation benchmarks/metrics like ROGUE and BLEU measure overlap of words, sequences, and symbols (n-grams) in order to provide a metric of summarisation (Davoodijam and Mohsen, 2024). It's probably difficult for ROGUE to distinguish between a given proposition and its negation—that is, removing a 'not' changes the meaning of the sentence quite heavily, but that might not affect how ROGUE

discussed in Section 2 are too ready to take their summaries at face value—or else to simply rely on ex post ratification of the summaries by the participants. If information processing tools are going to support democratic deliberation and decision-making, we should insist on high standards, in part because otherwise the conciliation function of democracy is undermined. Any given summary is likely to provide more support to one side or other of a competing set of interests. If the veracity of the summary can easily be challenged, then the disfavoured side has a ready-made reason not to accept the weight of this apparent evidence against them.[20]

## Aggregation

What about using LLMs to aggregate judgments, both to identify measures of support and to generate new statements that might better achieve consensus? This again seems to be a dangerously double-edged sword. The ability to support dialogues over policy development and to identify areas of potential common ground can clearly help the project of conciliation, by helping us find points of compromise where they didn't previously appear. But two concerns arise.

First, is public discourse on policy really held back by an incomplete universe of available alternatives? It is encouraging that these experiments show models can produce statements that command more attachment than the existing corpus, in these specific controlled settings, but we can't tell whether those statements constitute a significant *substantive* advance just from their being approved (perhaps they are simply vague enough to obscure disagreement). Moreover, the attachment they command is likely a function of the narrowly pro-social, apolitical setting of the conversations, and the low stakes of the policy debate in play. If the policy debate were more contentious and higher stakes, and if the participants were not independently motivated by the experimental setting to be open-minded and willing to compromise, we suspect these generated statements would be implicitly ideologically sorted (Bullock and Lenz, 2019) much faster, and would be less relevant to reaching a compromise between competing interests.

Put differently, in realistic political settings with a sufficiently open debate we can be reasonably confident that a wide enough range of policy proposals will be made.

---

scores models on their summarisations. And so perhaps the possibility of an extrinsic metric that pertains to the use of summarisations of democratic contexts should be researched (Davoodijam and Mohsen, 2024). In any case, we ought to be aware of what metrics are used to measure the efficacy of various LLMs at summarisation when choosing a given type of model to summarise in democratic contexts. We are indebted to an anonymous reviewer for pointing this out to us.

[20] Moreover, even after we decide to use an LLM for summarisation in a democratic decision procedure, we also need to pay attention to what *kind* of LLM we employ. For instance, GPT differs importantly from BERT-based models. While both are transformers, a big difference is that BERT analyses text *bidirectionally*, whereas GPT does so only from right to left. This makes BERT prima facie stronger at preserving context and at *extractive* summarisation—that is, summarisation that roughly aims at extracting words and sentences from linguistic data that representatively summarize the data (Batra et al., 2021; Bharathi Mohan et al., 2023; Darapaneni et al., 2023). Other LLMs like PEGASUS, for instance, are better at *abstractive* summarisation (Abdul Samad et al., 2024). It's also possible that GPT-based models—since they are more often used for generating text—might become better than BERT-based models at more abstractive summarisation that requires synthesising and generative capabilities (Darapaneni *et al.*, 2023). Indeed there are *many* architectural choices that can be made in the development of a summarisation tool—to know which choices better serve democratic values, we must start by reflecting on just what the democratic desiderata are for good summaries. We are indebted to an anonymous reviewer for pressing us on this point.

Democratic politics is probably not being held back by a lack of imagination. Whether compromise proposals will attract support is likely to depend on strategic facts about the nature of the competing interests from which these experiments all abstract away. As a result, using LLMs to aggregate judgments and suggest novel policy proposals is unlikely to do much to enhance real-world democratic decision-making and deliberation.

Second, there's a big difference between *generative* social choice and computational social choice more generally. When computational social choice theorists apply some aggregation function over a set of numerical preferences they are applying a transparent, mathematically verifiable algorithm to reach a decision (Brandt et al., 2016). Obviously the numerical representation of preferences is its own kind of artifice, but the decision procedure is reproducible and amenable to inspection and challenge. When the LLM produces some consensus statement based on its interpretation of the existing statements and people's preferences—and when the LLM is used to score similarity between statements—there is no explicit and deterministic aggregation algorithm being applied, the decision is entirely opaque and is likely not reproducible.[21] This is, of course, a long-standing and familiar problem with deep learning-based decision-making tools (Burrell, 2016; Selbst and Barocas, 2018; Vredenburgh, 2022; Lazar, 2024a). But it is an especially acute problem here with respect to the value of conciliation.

Suppose you are in the minority that is not favoured by the verdict delivered by the LLM. What reason do you have to accept that verdict? With computational social choice we can at least show the function that takes us from everyone's preferences to the outcome. And this is even simpler for a majoritarian voting procedure. But when the outcome is decided by an LLM, we cannot reproduce or inspect the decision-making function. Indeed, we cannot even guarantee that it wasn't materially affected by some exogenous influence. We cannot provide guarantees of security, or absence of the LLM's own bias. In higher stakes decision-making settings, there would be an obvious incentive to steer the mediating LLM by using adversarial techniques like prompt injection and jailbreaking (Greshake et al., 2023; Shah et al., 2023). The authors of 'Generative Social Choice' anticipate this concern somewhat by having participants ratify the model's restatements of theirs and others' views, but one might again think that if the stakes were higher people would be more discriminating (and in any case 16% of people saw some significant divergence between the generated statements and the inputs that led to it); moreover there might be some other slate of statements that they would endorse more.

## Representation

Turn next to the aspiration to use LLMs as representatives for individual voters.[22] Now, the most obvious objection here is one of performance—at present it is simply infeasible to design LLM representatives that reliably predict and track a principal's judgments. But let's set that concern aside for the moment, and try to evaluate the democratic case for these kinds of AI agents as proxies/representatives, if they work well. Our existing approaches to political representation already involve a principal-agent problem:

---

[21] A different random seed and temperature would likely lead to quite different results, for example.

[22] Compare here Lechterman, 2024, who considers the hypothetical possibility of deploying AI agents in the role of elected officials, negotiating with other officials to reach legislative compromises. See also Burgess, 2022, which takes a more favourable view of AI agents as politicians.

politicians often do a poor job of actually representing those who elect them (Besley, 2006). If we imagine AI agents along the lines envisaged in (Gudiño-Rosero *et al.*, 2024; Jarrett *et al.*, 2025), then we could surely make them more faithfully represent the preferences of their principal; if we could then come up with a fair decision rule for how to aggregate the judgments of those agents, we could potentially have a version of democratic politics that had some of the virtues of participatory democracy with no more and even less effort on citizens' part than representative democracy. At least, the experiment is worth running.

But we should proceed with eyes open. Any use of AI agents to represent voters must apply a decision rule of some kind, and as just argued it's hard to see how those whom the decision disfavours could be reasonably expected to accept those decisions. This approach also obviously undermines the *participatory* reasons to value democracy. One cannot develop civic virtue by delegating politics to an AI agent. And it can also undermine the democratic value of *preference transformation* and collective preference formation. Projects that rely extensively on training LLMs to represent the participants (Gudiño-Rosero *et al.*, 2024; Fish *et al.*, 2025; Jarrett *et al.*, 2025; more deliberative projects like Recursive Public might avoid these objections) fail on the following two counts.

First, they treat the participants' preferences as fundamentally static. They extrapolate from their preferences in an initial survey, without giving them the opportunity to update their positions. This problem could be resolved by adopting some kind of multi-turn design. The second problem is more fundamental. These approaches cannot provide participants opportunities to change their preferences as a result of engaging with others with whom they disagree. This kind of interaction is valuable not only because it yields outcomes that represents participants' better-considered preferences, but because engaging with the other directly, seeing the matter under contention from their point of view, and coming to adjust your preferences through a process that you associate with that exchange, is valuable in its own right. It serves both the value of conciliation, and that of preference transformation, in particular insofar as it leads to the formation or identification of a general will (Rousseau, 1762; Young, 2002).

We are also sceptical about whether delegation to proxy agents would preserve the epistemic benefits of democracy—for example, how would the Condorcet jury theorem be affected by deputising Language Model Agents in place of citizens? Would they really have the same cognitive diversity as the people they putatively represent (Landemore, 2012a, 2021a)?

In fact, one can argue that an ersatz deliberation among AI agents would omit some of the key mechanisms by which democratic politics leads to epistemically better decisions. For example, our human adversarial political process helps surface information relevant to collective decision-making that would otherwise remain submerged. And it's only because they actively participate in public life that political visionaries come up with novel policy proposals that don't just triangulate among the things that people already believe, but offer genuinely innovative solutions to pressing practical problems. In other words, democratic procedures (1) pool knowledge that is normally distributed (2) force different kinds of knowledge to interact with each other, leading to corrections, the surfacing of more reasons and proposals, and so on (Landemore, 2012b), and (3) provide opportunities for creative leadership that exceeds what is feasible by simply operating over pre-existing preferences. Perhaps we could design a system for representative AI agents that also has these virtues, but it would require much more than just training capable deputies; it would need the creation of a virtual political sphere to enable their interaction (this presents more exciting opportunities for computational political theorists).[23]

### Facilitation

The facilitative role of LLMs in online deliberation processes seems on the whole much more benign (though for some scepticism see Alnemr, 2020; Malkin and Alnemr, 2024). On the other hand, the availability of human moderators and facilitators does not seem as important a bottleneck for deliberative democracy as is actual voters' unwillingness to participate in intensive, time-consuming deliberation. Moreover, we suspect that much of the benefit of successful deliberative democratic procedures depends on the fact that they are held in person, and involve people facing and seeing one another as whole people, to whom they are to some degree accountable.[24] The appetite for reaching meaningful consensus or making progress towards it is likely much greater for in-person convenings than for virtual ones (Hartz-Karp and Sullivan, 2014). Put another way, successful democratic procedures might necessarily be embodied. So the relative success of LLMs at facilitating some kinds of online deliberation might be only a modest count in their favour.

## 5. How Can LLMs Help Democracies?

The projects under discussion narrowly focus on the question of whether LLMs can be used to support the participatory design of chatbots and the norms that govern them. Rightly, their authors were not trying to 'fix' democracy. However, some clearly do think that LLMs have the potential to significantly transform and reinvigorate democratic institutions, and the foregoing critique should offer some grounds for caution (Landemore, 2023; Ovadya, 2023). In particular, we want to highlight three broad lessons about whether and how LLMs can support the instrumental goals of democracy, which we think can offer an instructive practical guide to those who are tempted to build on these initial results in more ambitious ways.

Democracies aim to realise many different goals, but perhaps foremost among them is to enable people with often radically conflicting values and directly competing interests to live together in relative harmony. LLMs are simply the wrong kind of tool to reconcile competing interests and values. Procedures for that purpose need to be fair, simple, secure, transparent and contestable. A majority vote, with some constitutional constraints to protect persistent minorities and basic rights, meets those criteria. Any attempt to use AI to support fair procedures for resolving competing claims must take these criteria as constraints. LLMs in particular, and deep neural networks in general, cannot at present satisfy these criteria and arguably will never be able to do so.

---

[23] One possible avenue for development is multi-agent frameworks. Frameworks that deploy multiple LLM agents have yielded promising results on a number of metrics (Chan et al., 2023; Wu et al., 2023). That being said, even multi-agent frameworks using state-of-the-art LLMs still underperform humans on complex tasks that require coordination (Abdelnabi et al., 2024). Considering that deliberation and other forms of democratic decision-making are exactly such tasks, we don't think multi-agent frameworks currently offer the kind of technical ability needed to instantiate the epistemic virtues that can justify democratic governance.
[24] This is contentious; some proponents of online deliberation think that we can realise as good results from deliberation online as in person.

Despite some promising attempts to use LLMs to facilitate deliberation, they are unlikely, in the end, to contribute much to the process of preference transformation whereby polities aim at more than just a modus vivendi or transactional compromise, and seek to resolve their disagreements by arriving at something like a general will. Real democratic institutions must operate in a world riven by background inequality of power and resources, as well as deep ideological disagreement, where citizens are not able to avoid pressure and constraint (and in particular where partisan groups dominate political discussion), where decisions affect their material and cultural interests very closely, and where competing interests and values place people in an adversarial relationship with one another (Lovett, 2024). By contrast, each of the discussed projects places participants in a privileged, jury-like position where they engage in unfettered, unmanipulated, dispassionate debate over a set of problems that do not, for the most part, directly concern them. They are as close to the Habermasian 'ideal speech situation' as is practically feasible (Habermas, 1984). In these advantageous conditions, we suspect that *any* methods of democratic deliberation and decision-making would look attractive. LLM-supported deliberation might work well, but frankly traditional representative democracy would too if we bracketed all the contentious elements of politics and economics. Under realistic conditions, however, LLM-supported democratic deliberation is no more likely to make a difference than other versions of deliberative democracy. Moreover, insofar as using LLMs for facilitation steers us towards an online-only version of deliberation, we think that it foregoes one of the key virtues of deliberation, namely that by bringing people together, face to face, it does induce some willingness to give ground and reach compromise—we suspect, as argued above, that this will only robustly happen through in-person interaction.

However, while we caution against using LLMs to replace fair decision procedures, or to facilitate democratic deliberation, we think they can still make important contributions to renewing democracy. While we do not think that integrating LLMs into the formal public sphere of executive decision-making is likely to advance the epistemic aims of democracy, we *do* think that deploying LLMs in the *informal* public sphere can help provide some of the epistemic preconditions for democracies to flourish. The informal public sphere is the environment of unstructured dialogue and deliberation among members of a political community (Habermas, 1998; Young, 2002). The formal public sphere is the space of officially sanctioned deliberation and decision-making, in which concrete institutional action is effectively debated and taken[25]

Healthy democracies rely on a healthy informal public sphere to hold the powerful to account and to set a positive forward direction for society by facilitating informal collective deliberation (Habermas, 1998; Young, 2002; Cohen and Fung, 2021; Lazar, 2023). Our existing digital infrastructure for the public sphere is doing a bad job (Cohen and Fung, 2021). LLMs to date have only harmed the public sphere through the generation of search engine-optimised slop that supercharges the work of content farms. Their positive benefits in this domain are only just being fully explored.

[25] Obviously there is no bright line between formal and informal public spheres. We simply advocate that given the choice between interventions that enhance informal processes that contribute to democratic culture, and those that propose to replace key institutions for decision-making and action, AI researchers should err towards the former.

Early work suggests that frontier LLMs can be used to radically improve automated approaches to content moderation; they have also been shown to be excellent at reranking social media feeds in line with broader societal values (Bernstein et al., 2023; Jia et al., 2023). LLMs can probably offer a new way to use the internet in general, serving as an alternative to existing recommender systems that contribute to societal dysfunction, enabling us to navigate the digital public sphere in ways that discourage polarisation and extremism, and spare us from surveillance and manipulation (for a proof of concept see Jia *et al.*, 2023, for a comprehensive moral defence see Lazar et al., 2024; and for an overview of the technical literature see Lin et al., 2023; Vats et al., 2024). Obviously no technological solution is going to fix the public sphere, but a novel technique for filtering and recommending content opens up valuable new possibilities for amelioration. And LLMs' information retrieval capabilities and their capacity to either translate or simplify text and speech could provide valuable democratic resources for citizens who struggle to speak the national language, or who need complex personalised advice that they cannot effectively get from just searching government websites. And beyond just filtering, recommendation, and retrieval, LLMs have shown promising results in combatting beliefs in conspiracy theories with consistent, tailored counterarguments, and at facilitating more measured dialogue online (Rosenblum and Muirhead, 2019; Argyle *et al.*, 2023; Papaioannou et al., 2023; Costello et al., 2024).

We also think that some of the deliberative forums defended in these projects could offer an additional useful contribution to the public sphere. Recursive Public's iterative forum for deliberation and the Democratic Policy Development project could provide democratic citizens with novel ways of interacting with their co-citizens, discovering what they believe and value, and forming their preferences and values together. Provided their role is confined to this informal setting, without affecting to offer a procedure for settling disagreements and so on, this could be a useful addition to the toolkit of the democratic society.

Of course, just as with summarisation above, the fact that democracies would benefit from functional LLMs performing this role does not entail that they can do so *now*. Today's LLMs cannot be trusted in these roles. However, these are roles for which LLMs can be meaningfully trained and evaluated. With the right kind of collaborative approach—integrating AI science with political theory—these limitations are very likely to be surmountable, resulting in genuine democratic benefits. We think that this kind of careful collaboration could surmount some of our objections to democratic uses of LLMs described above. In particular, with careful reflection on and precisification of the criteria for appropriate summarisation of responses to calls for public input, we think LLMs definitely *could* enable much more effective extraction of public insights on key matters of state.

## 6. Conclusion

Recent advances in AI have been unprecedented—whatever the shortcomings of large language models, they constitute a radical expansion of the technological horizon. Saliency bias alone would be sufficient to ensure that LLMs' prospective usefulness for democratic deliberation and decision-making would be explored; but there is clearly more potential here than just that. Democracy is a fundamentally communicative practice, and it is therefore also a linguistic practice. New tools for analysing and producing linguistic content are sure to be useful at some point in the democratic pipeline. However, communication is ultimately not reducible to language alone—democracy is also fundamentally social. It is a way of surfacing and attending to interests, reaching compromise, and engaging in a kind of transformative experience (Paul, 2014). While there is little harm in using LLMs to support innovative approaches to the participatory design of chatbots, these experiments cannot offer much inspiration for broader democratic innovation. LLMs are too opaque to provide a legitimate democratic decision procedure. And we cannot outsource the social aspects of democratic communication to language models—we have participatory reasons that we ourselves should engage. Perhaps more can be said about the idea of having AI agents act as our political representatives (Burgess, 2022; Lechterman, 2024), but on first principles it seems a bad idea.

LLMs are not going to fix democracy. However, they can potentially foster healthier democracies—not by introducing novel methods of judgment aggregation or algorithmic representation, but by helping us better navigate the information glut that we face online. A healthy informal public sphere is a necessary precondition for a healthy democracy. LLMs' contribution to the former has so far been decidedly equivocal; but not only is there room for improvement, there is real potential to do so. We encourage AI scientists motivated to use LLMs to enhance democracy—and computational political theorists more generally—to shift their focus away from attempts to build alternative decision procedures to resolve competing interests and values, and from related efforts to supplement formal forms of democratic deliberation, and to focus instead on using the information retrieval and presentation capabilities of LLMs to give democratic citizens the wherewithal to navigate and improve our informal public sphere.